\documentclass[letterpaper, english, aps, prx, reprint, superscriptaddress, longbibliography]{revtex4-2}
\pdfoutput=1
\usepackage{bbm}
\usepackage{color}
\usepackage{xcolor}
\usepackage{pdfcolmk}
\usepackage{verbatim}
\usepackage{mathtools}
\usepackage{amsmath}
\usepackage{amssymb}
\usepackage{float}
\usepackage{array}
\usepackage{mathrsfs}
\usepackage{graphicx}% Include figure files
\usepackage{dcolumn}% Align table columns on decimal point
\usepackage{bm}
\usepackage{bibentry} 
\usepackage{amsmath,amsthm}
\usepackage{amssymb,amsfonts}
\usepackage{bbm}
\usepackage{graphicx}
\usepackage{subfigure}
\usepackage{afterpage}
\usepackage{times}
\allowdisplaybreaks[1]
\usepackage[mathcal]{euscript}

\newtheorem*{ansatz*}{Ansatz}

\newcommand{\be}{\begin{equation}}
\newcommand{\ee}{\end{equation}}

\newcommand{\bse}{\begin{subequations}}
\newcommand{\ese}{\end{subequations}}
\newcommand{\ket}[1]{\left|{#1}\right\rangle}

\newcommand{\Z}{\mathbb{Z}}

\newcommand{\bra}[1]{\langle{#1}|}

\newcommand{\bpm}{\begin{pmatrix}}
\newcommand{\epm}{\end{pmatrix}}
\newcommand{\bmm}{\begin{matrix}}
\newcommand{\emm}{\end{matrix}}% bold math
\makeatletter

\providecolor{lyxadded}{rgb}{0,0,1}
\providecolor{lyxdeleted}{rgb}{1,0,0}
\DeclareRobustCommand{\lyxsout}[1]{\ifx\\#1\else\sout{#1}\fi}

\usepackage[colorlinks,citecolor=blue,linkcolor=blue]{hyperref}

\begin{document}
%\bstctlcite{apsrevcontrol}

\title{Simulating non-Abelian statistics of parafermions with superconducting processor}

\author{Hong-Yu Wang}
%\thanks{First author}
\affiliation{International Center for Quantum Materials and School of Physics, Peking University, Beijing 100871, China}
% \affiliation{Collaborative Innovation Center of Quantum Matter, Beijing 100871, China}
\affiliation{International Quantum Academy, Shenzhen 518048, China}
\affiliation{Hefei National Laboratory, Hefei 230088, China}
\author{Xiong-Jun Liu}
\thanks{Corresponding author: xiongjunliu@pku.edu.cn}
\affiliation{International Center for Quantum Materials and School of Physics, Peking University, Beijing 100871, China}
% \affiliation{Collaborative Innovation Center of Quantum Matter, Beijing 100871, China}
\affiliation{International Quantum Academy, Shenzhen 518048, China}
\affiliation{Hefei National Laboratory, Hefei 230088, China}

\begin{abstract}
Parafermions, which can be viewed as a fractionalized version of Majorana modes, exhibit profound non-Abelian statistics and emerge in topologically ordered systems, while their realization in experiment has been challenging. Here we propose a novel experimental scheme for the quantum simulation of parafermions and their non-Abelian braiding statistics in superconducting (SC) circuits by realizing the $\mathbb{Z}_d$ plaquette model on a two-dimensional lattice. Two protocols using quantum circuits and non-destructive measurements are introduced to prepare the topologically ordered ground state, on which parafermion pairs are created by engineering dislocations. We then propose a generalized code deformation approach to realize the fusion and non-Abelian braiding of parafermions, and show the application of this approach to the $\mathbb{Z}_3$ parafermions. We also examine the real experimental parameter regime to confirm the feasibility of our scheme in SC devices. This work extends previous quantum simulations of topological defects in SC qubits to qudit systems and opens up a promising way for parafermion-based high-dimensional topological quantum computing with experimental feasibility.
\end{abstract}

\maketitle

\section{Introduction}\label{sec:intro}
The rapid advancement of quantum processors has enabled the simulation of many-body quantum states and the exploration of intriguing quantum phenomena~\cite{Seth1996,Lanyon2011,Barends2015,Smith2016,Martinez2016,Zhang2017,Bernien2017,Chiesa2019,de2019observation,Motta2020,Monroe2021,Satzinger2021,Semeghini2021,Mi2021,Atas2021,Mi2022,Zhang2022,Dumitrescu2022,Noel2022,Wang20222,Smith2022,Graham2022,Chen2024,Dhruv2024}. An particularly exciting direction is the simulation of anyons~\cite{Kiczynski2022,Mi2023,Dvir2023,Bordin2024,tenHaaf2024,tenHaaf2025}, exotic quasiparticles emerging in topologically ordered states (topological orders)~\cite{Wen1989,Wen1990}, whose realization and manipulation remain challenging in materials~\cite{HANBURY1956,Lu2009,Rosenow2016,Nakamura2020,Bartolomei2020,Nakamura2022}. Recent digital simulations using superconducting (SC) qubits have demonstrated non-Abelian braiding statistics of engineered topological defects in the $\mathbb{Z}_2$ toric code model~\cite{Kitaev2003,Bombin2010,Fowler2012,Xu2023,Andersen2023,Fauseweh2024}. These defects, associated with one-dimensional lattice dislocations, exhibit fusion and projective braiding properties identical to those of intrinsic point-like non-Abelian anyonic excitations~\cite{Barkeshli2013}, enabling the global encoding of quantum information~\cite{Elfeky2021}. Consequently, braiding these topological defects to process quantum information is topologically protected at the logical level~\cite{Song2018}. This protection, rooted in the topological nature of the surface code model and the non-local encoding with topological defects, offers a promising pathway toward fault-tolerant quantum computing~\cite{Liu2022,Fan2022ohd,Xu2023,Andersen2023,Xu2024Fib,Iqbal2024,Goel2024,Zatelli2024}.

Topological defects in the $\mathbb{Z}_2$ toric code model manifest the non-Abelian statistics of Majorana fermion modes, which is described by the Ising anyon model~\cite{Bombin2010,Kitaev20062}. Quantum computing schemes based on Majorana modes have been widely studied~\cite{kitaev2001unpaired,ivanov2001non,alicea2011non,alicea2012new,aasen2016milestones,mourik2012signatures,deng2012anomalous,rokhinson2012fractional,das2012zero,Kraus2013,wang2012coexistence,churchill2013superconductor,xu2014artificial,nadj2014observation,chang2015hard,Sarma2015,albrecht2016exponential,wiedenmann2016,Schrade2018,zhang2018observation,wang2018evidence,fornieri2019evidence,ren2019topological,jack2019observation,Aguado2020,aghaee2023inas,Hodge2025,Aghaee2025}, but their experimental realization in one- or two-dimensional condensed matter systems faces significant challenges~\cite{takei2013soft,liu2017phenomenology,bocquillon2017gapless,ahn2021estimating,pan2021disorder,yu2021non,knapp2020fragility,Woods2021,Hess2023,Kouwenhoven2025}. More critically, braiding Majorana modes alone is insufficient for universal quantum computing~\cite{Kitaev20032,Nayak2008}. Achieving universality requires additional non-topological assistance methods, which compromise the inherent topological protection~\cite{Bravyi2005,Ahlbrecht2009,Ezawa2024}. $\mathbb{Z}_d$ parafermions generalize Majorana modes (which are a special case for $d=2$) and exhibit more complex and nonlinear non-Abelian braiding statistics~\cite{Benhemou2023,Levaillant2015}. In contrast to Majorana modes, parafermions can encode multi-level computational units, known as qudits, which provide a larger state space for storing and processing information~\cite{Matthew2009,Chi2022,Ringbauer2022}. This enables the reduction in circuit complexity and enhancement in algorithmic efficiency~\cite{Wang2020}. High-dimensional quantum computing with qudits also offers further advantages, such as greater flexibility in simulating quantum dynamics~\cite{Blok2021,Matthew2009} and higher-fidelity magic state distillation~\cite{Anwar2012,Campbell2012}. Unlike Majorana modes, braiding of parafermions enables entanglement gates. In particular, it allows the generation of the single-qudit Clifford group for any $d$ and the multi-qudit Clifford group for odd $d$~\cite{Hutter2016}. Non-Clifford gates can be implemented by incorporating flux braiding through the Aharonov–Casher effect~\cite{Dua2019}, complementing parafermion braiding to realize a universal gate set. Moreover, parafermions can serve as building blocks for engineering Fibonacci anyons~\cite{Vaezi2014,Mong2014}, whose braiding supports universal quantum computation. As such, $\mathbb{Z}_d$ parafermions with $d > 2$ are computationally more powerful than Majorana modes, bringing us closer to universal quantum computing.

Proposals for realizing parafermions in condensed matter systems—such as fractional quantum Hall states or fractional topological insulators with induced superconductivity, and coupled nanowires—typically require strong interactions or fractionalization, which however pose significant experimental challenges~\cite{Lindner2012,clarke2012parafermion,Cheng2012,Clarke2013,Vaezi2013,Klinovaja20143,Klinovaja2014,Klinovaja20142,Cobanera2015,Alicea2016,Ebisu2017,Laubscher2019,Laubscher2020,Santos2020,Schiller2020,Khanna2022,Teixeira2022,Hong2024}. Alternatively, engineering topological defects in Abelian topological orders provides a general route to realizing non-Abelian statistics~\cite{Barkeshli2013}, which can extend beyond the Majorana case. Since Abelian topological orders are more prevalent and generally easier to realize than their non-Abelian counterparts, this approach offers a simpler and more experimentally feasible path to realizing parafermions. Additionally, existing Abelian anyonic excitations facilitate the implementation of phase gates. Following this idea, topological defects obeying parafermion statistics are theoretically predicted in a stabilizer Hamiltonian model known as the $\Z_d$ plaquette model~\cite{You2012}, though experimental realizations remain limited to the simplest case of Majorana modes ($d=2$)~\cite{Dai2017,Xu2023,Andersen2023}. Digital quantum simulation of the $\mathbb{Z}_d$ plaquette model with topological defects presents a promising approach for realizing parafermions, enabling state preparation and evolution via quantum circuits without relying on microscopic Hamiltonians~\cite{You2012,You2013,Mei2020,Lensky2023,Iqbal2024}. Well-separated topological defects can globally encode logical qudits, functioning similarly to point-like intrinsic parafermionic excitations~\cite{Bombin2010,Cong2017}. In practical implementations, an error correction strategy is needed to manage stray anyons~\cite{Wootton2014}. Fortunately, parafermions realized through defects associated with extended structures exhibit enhanced robustness due to their extrinsic nature (i.e., they are not part of the spectrum of the underlying Abelian model) and limited mobility~\cite{Iris2017,Wang2022}. As a result, non-Abelian error correction is not required; standard surface code error correction on the background Abelian model is sufficient to ensure fault tolerance~\cite{Hutter2015}. In surface code error correction, topological protection is characterized by the code distance—the number of local operations required to transform one ground state into another. For the $\mathbb{Z}_d$ plaquette model with parafermion defects positioned at the center of a planar surface, the code distance grows linearly with system size, resulting in an exponential suppression of logical qudit errors. This approach is particularly relevant for noisy intermediate-scale quantum (NISQ) devices~\cite{Preskill2018}, offering valuable insights into fault-tolerant, high-dimensional quantum information processing.
%Quantum circuits composed of local unitary operators can manipulate these defects to realize braiding. While circuits may encounter errors, the system ensures quantum information processing against errors that do not alter the topology of the braiding worldlines. \

Simulating the $\Z_d$ plaquette model with parafermion defects requires precise coherent control of qudits. SC devices, such as transmons with readily addressable higher energy levels, offer a promising platform for this purpose~\cite{Blais2021}. Recent advancements have demonstrated high-fidelity single- and two-qutrit (qudit with $d=3$) entangling operations ~\cite{Cao2024,Roy2023,Luo2023,Yurtalan2020,Morvan2021,Goss2022,Goss2024}. Nevertheless, realization of the parafermion modes and implementation of their non-Abelian statistics necessitate qualitatively more complicated quantum gate operators and involve addressing non-trivial physical challenges in comparison with the Majorana modes~\cite{Andersen2023,Xu2023}, and it remains unclear how to achieve such quantum simulation in the real SC circuits. While the mathematical extension from Majorana to parafermion modes may seem straightforward, physical realization is considerably more intricate. First, the fundamental single- and two-qudit gates are unitary but not Hermitian, requiring careful consideration of both the operations and their Hermitian conjugates in circuit implementations. Second, the multi-level structure of qudits necessitates novel physical processes to control level transitions and to implement entangling operations involving higher levels. Examples include the two-photon process for controlling the $0$-$2$ transition in qutrits~\cite{Yurtalan2020} to realize high-fidelity generalized Hadamard gate, and implementing control gates based on tunable cross-Kerr interactions~\cite{Luo2023,Goss2022}. Third, parafermions exhibit much richer non-Abelian braiding statistics and fusion rules compared to Majorana modes, making it physically non-trivial to identify suitable observables for characterizing their braiding and fusion properties. This work aims to address these challenges and integrate recently developed qudit techniques in SC devices to develop an experimentally feasible scheme for simulating parafermions and their fusion and braiding.

Here, we summarize the main results of our work. In this work, we propose an experimentally feasible scheme to simulate parafermions and their fusion and non-Abelian braiding in SC circuits, focusing on the $\mathbb{Z}_d$ plaquette model on a planar surface. We introduce two efficient ground state preparation protocols: one using quantum circuits and the other employing non-destructive measurements. By engineering dislocations in the lattice, pairs of parafermions are created. With a novel generalized code deformation approach, as proposed in the present work, these parafermions are moved on the planar surface, enabling a feasible way for their fusion and braiding operations.
The concrete example of $d=3$ parafermions is studied in detail. The real parameter conditions are examined, with the feasibility of our scheme in experiment being confirmed. We note that while the present simulation is designed for the SC devices, it is also compatible with other experimental platforms capable of implementing the gates that we use. Our work provides a theoretical foundation for the imminent quantum simulation of parafermions in quantum computing platforms and advances topologically protected high-dimensional quantum computing.

The remainder of the paper is organized as follows. In Sec.~\ref{sec:model}, we review the $\mathbb{Z}_d$ plaquette model, its elementary excitations, and parafermion defects. In Sec.~\ref{sec:gscpara}, we present the protocol for ground state preparation and parafermion creation. Sections~\ref{sec:fusion} and~\ref{sec:braiding} outline the schemes for realizing the fusion and non-Abelian braiding of parafermions. In Sec.~\ref{sec:experpara},  we discuss the experimental parameters of the SC devices for realizing our scheme. Finally, in Sec.~\ref{sec:outlook}, we summarize our findings and discuss potential applications in high-dimensional topological quantum computing.

\section{The Model}\label{sec:model}

\begin{figure*}
    \centering
    \includegraphics[scale=0.24]{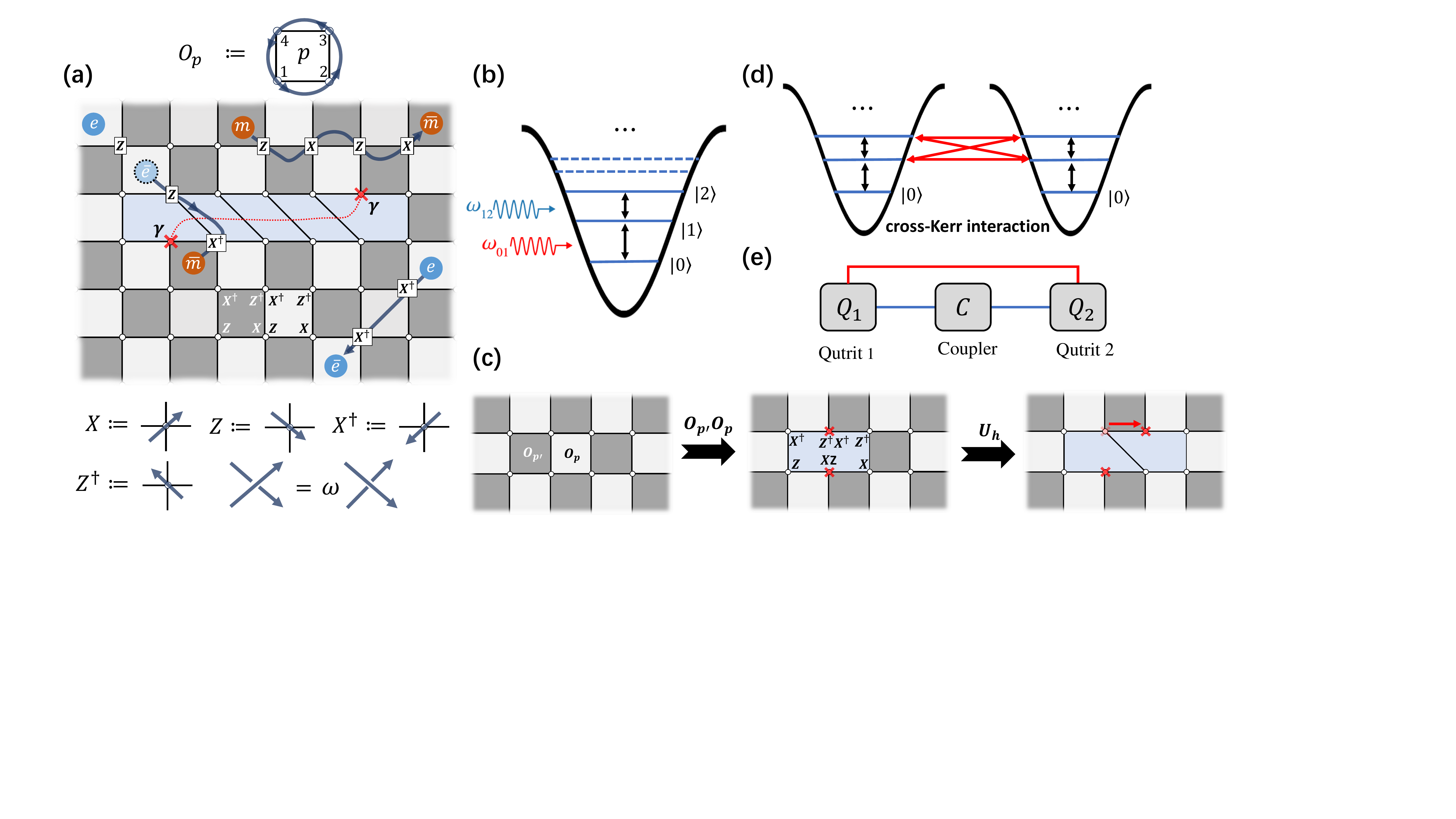}
    \caption{Illustration of the model with a pair of prafermions. (a) The $\mathbb{Z}_d$ plaquette model on a checkerboard lattice with a dislocation colored by light blue, where the small white dots represent qudits. The red cross marks the defect associated with the dislocation, identified as the parafermion mode $\gamma$. The red dotted line indicates the branch cut connecting the two parafermions. The arrowed dark blue curves represent string operators that create $e-\overline e$ and $m-\overline m$ excitations, which reside on the light and dark plaquettes, respectively. In our convention, the $e$ or $m$ particle is located at the start of the string operator and its dual is located at the end of the string operator. Stabilizers $O_p$ on both dark and light plaquettes are shown. Graphical representations of the generalized Pauli operators, commutation relation, and stabilizers are shown. (b) An illustration of the energy diagram and microwave driving of a typical transmon as a multilevel system. The frequency difference $\omega_{12} - \omega_{01}$ characterizes the anharmonicity. (c) Dislocations are engineered with two neighboring stabilizers being replaced with a single one. Applying a certain two-qudit quantum circuit moves and separates the parafermion defects. (d) Schematic of the cross-Kerr interaction that is used to realize two-qutrit gates. (e) A  configuration that realizes the tunable cross-Kerr interaction using a frequency-tunable coupler, as demonstrated in Ref.~\cite{Luo2023}.}
    \label{fig:figureone}
\end{figure*}

We start by considering the $\mathbb{Z}_d$ plaquette model on the two-dimensional square lattice with a checkerboard pattern [Fig.~\ref{fig:figureone}(a)]. On each vertex $i$, there is a qudit with $d$ basis states $\ket{m_i}$ ($m_i=0,1,2...,d-1$). Physically, the qudit can be realized by the intrinsic multi-dimensional Hilbert space of SC devices, such as transmons. Their coherence properties and sufficient anharmonicity enable to function as qudits, with the lowest $d$ energy levels serving as the qudit basis states [Fig.~\ref{fig:figureone}(b)].  The Hilbert space is spanned by the qudit states on all the vertices. The Hamiltonian is:
\begin{equation}\label{eq:ham}
 H = -\sum_{p} O_p + h.c., \, O_p =  Z_1 X_2 Z_3^{\dagger}X_4^{\dagger},
\end{equation}
where the labels of the qudits are shown in the top of Fig.~\ref{fig:figureone}(a), and the sum of plaquettes $p$ is for all the light and dark plaquettes. Here, $Z$, $X$, $Z^{\dagger}$, and $X^{\dagger}$ are generalized Pauli operators for qudits. They are defined by the properties: $X^d = \mathbbm{1}, \, Z^d = \mathbbm{1}, \,  XZ = \omega ZX,$ where $\omega = e^{\frac{2\pi i}{d}}$ and act on the qudit basis as $Z \ket{m} = \omega^m \ket{m}$ and $X \ket{m} = \ket{m-1 \, (\text{mod} \quad d)}$. The experimental realization of single-qudit operations can be achieved through microwave driving resonant with specific qudit state levels. For $d=3$ (i.e., a qutrit), any single-qutrit gate can be decomposed into a sequence of operations acting on two-level subspaces~\cite{Blok2021,Morvan2021}, and implemented with high fidelity by accounting for the AC Stark effect arising from anharmonicity~\cite{Blok2021,Liu2016}. It is convenient to introduce a graphical representation [Fig.~\ref{fig:figureone}(a)] for the generalized Pauli operators and their commutation relations. With this representation, the stabilizer $O_p$ is described by a loop operator around the plaquette $p$, which satisfies $O_p^d = 1$ and has eigenvalues $1$, $\omega$,..., $\omega^{d-1}$. Stabilizers on different plaquettes commute with each other; therefore, the Hamiltonian Eq.~\eqref{eq:ham} is exactly solvable. The ground state $\ket{\Psi_0}$ satisfies:
\begin{equation}\label{eq:gs}
 O_p\ket{\Psi_0} = O_p^{\dagger}\ket{\Psi_0} = \ket{\Psi_0}, \, \forall p,
\end{equation}
and the ground-state projector is:
\begin{equation}\label{eq:gspro}
 \Pi^{0} = \prod_{p}\Pi_p^{0}, \quad \text{where} \quad \Pi_p^{0}=\frac{1}{3}\left(\mathbbm{1}+O_p+O^{\dagger}_p\right).
\end{equation}
The model defined by Eq.~\eqref{eq:ham} provides a Hamiltonian realization of the topological gauge theory with gauge group $\mathbb{Z}_d$. The stabilizers on light (dark) plaquettes correspond to local Gauss’s law (flatness condition) constraints. The ground state, defined by Eq.~\eqref{eq:gs}, is the state where all local constraints are satisfied, while excited states arise from violations of these constraints. For $d=2$, the stabilizer has only two eigenvalues $1$ and $-1$, indicating whether a local constraint is satisfied or violated. The elementary excitations—violations of Gauss’s law and flatness condition—are termed charge ($e$) and flux ($m$) excitations, respectively, following its physical meaning in gauge theory. In the $\mathbb{Z}_2$ gauge theory, these excitations are self-dual, satisfying $e^2=1$ and $m^2=1$. For $d\geq 3$, however, charge and flux excitations are no longer self-dual and can appear in multiple powers. Specifically, the elementary excitations of Eq.~\eqref{eq:ham} include charge excitations $e, e^2,...,e^{d-1}$ on light plaquettes and flux excitations $m,m^2,...,m^{d-1}$ on dark plaquettes (see Appendix.~\ref{sec:elementarryex} for an algebraic characterization of these excitations).  More general excitations involve composites $e^{k}m^{d-k}$, known as dyons. When a charge  excitation $e$ encircles a flux excitation $m$ counterclockwise, or vice versa, the system acquires a phase factor $\overline{\omega}\equiv \omega^{d-1}$ due to the Aharonov-Bohm effect. Consequently, charge, flux, and dyon excitations are Abelian anyons, created by string operators [graphically represented in Fig.~\ref{fig:figureone}(a)]. The ground state of Eq.~\eqref{eq:ham} exhibits the Abelian topological order characterized by the quantum double $D(\mathbb{Z}_d)$~\cite{Kitaev2003}.

To incorporate non-Abelian statistics with the Abelian topological order model, we need to engineer dislocations. The dislocation in the square lattice that we consider is an extended object formed by tilting the edges along a line [the light blue region in Fig.~\ref{fig:figureone}(a)]. Associated with the dislocation are two point-like defects characterized by trivalent vertices on the pentagon plaquettes. In the $\mathbb{Z}_d$ plaquette model, the dislocation is associated with deforming stabilizers $O_p$ and $O_{p'}$ on two neighbouring plaquettes to a single stabilizer $O_{p'}O_{p}$ [Fig.~\ref{fig:figureone}(c)]. This deformation effectively removes the edge between the plaquettes and reduces the number of stabilizers by one. Therefore, $N$ such dislocations result in the $Nd$-fold degeneracy, with each defect having a quantum dimension $\sqrt{d}$ that indicates the non-Abelian nature of the defect. These non-Abelian defects manifest parafermionic behavior, whose underlying physics is the defect-induced electric-magnetic exchange symmetry. To illustrate this, consider creating an $e-\overline e$ charge pair by applying a $Z$ gate to a qudit, and moving the $\overline e$ particle through the branch cut connecting two defects using a string operator [Fig.~\ref{fig:figureone}(a)]. The $\overline e$ excitation transforms into the $\overline m$ excitation, indicating an $e-m$ exchange. This shows that the defect identifies the dyon exictation $e^k m^{d-k}$ ($k=1,2,...,d-1$) with the trivial topological charge $1$, giving rise to the following fusion rules:
\begin{equation}\label{eq:fusionrule}
    \gamma \times \gamma = 1 + \sum_{k=1}^{d-1} e^k m^{d-k},
\end{equation}
where $\gamma$ denotes the defects. Equation \eqref{eq:fusionrule} is exactly the fusion rules of parafermions. For clarity, we henceforth refer to these defects as parafermions. By applying certain two-qudit operations, whose circuit realizations are shown later, we can move the parafermions apart in space. This allows to realize the fusion and braiding of parafermions. For convenience, we focus on the qutrit case, i.e., the $\mathbb{Z}_3$ plaquette model in the following. Generalizing the methods to larger $d$ is straightforward.

\section{Ground state preparation and creation of parafermions}\label{sec:gscpara}
\begin{figure*}
    \centering
    \includegraphics[scale=0.25]{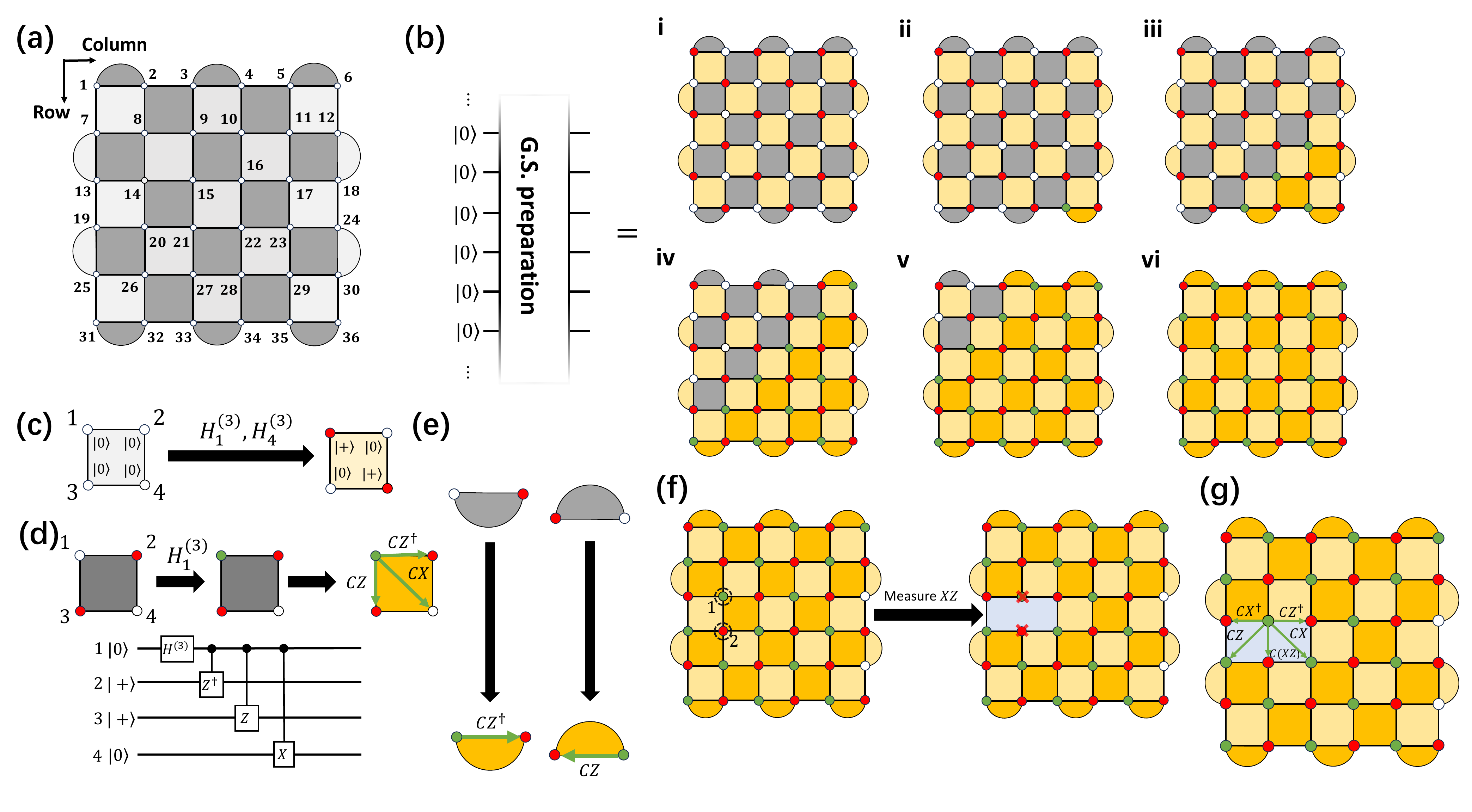}
    \caption{Illustration of the ground state preparation and parafermion creation protocol. (a) The $\mathbb{Z}_3$ plaquette model with a $6\times 6$ array of qutrits. (b) Explicit procudure for the ground state preparation. The system is initialized with all qutrits in $\ket{0}$. Red dots indicate the representative qutrits used to prepare the light plaquettes in the ground state, while green dots represent the control qutrits used to apply controlled two-qutrit gates for preparing the dark plaquettes in the ground state. The plaquettes in the ground state are colored pale yellow for light plaquettes and gold for dark plaquettes. (c) Applying the generalized Hadamard gate to the representative qutrits prepares the light plaquettes in the ground state. (d) A circuit is applied to the dark plaquettes to prepare them in the ground state. (e) Circuits for preparing boundary dark plauettes in the ground state. (f) Measuring two neighboring qutrits (encircled by dashed line) creates a pair of parafermions. (g) An alternative method for creating parafermions involves ground state preparation on a lattice with a hole, represented by a light blue region.}
    \label{fig:figureTwo}
\end{figure*}

In this section, we present methods to efficiently prepare the ground state of the stabilizer Hamiltonian in Eq.~\eqref{eq:ham} on a planar surface. With the ground state, we show how parafermions can be created. For illustration, we focus on the case of $d=3$, where the fundamental degrees of freedom are qutrits.
\subsection{Ground state preparation}
We consider the $\mathbb{Z}_3$ plaquette model with $6\times 6$ qutrits [Fig.~\ref{fig:figureTwo}(a)]. The system is initially in the product state $\ket{0}^{\otimes 36}$. The ground state of the model is long-range entangled, which requires two-qutrit entangled operations for the preparation. This can be experimentally achieved through the tunable cross-Kerr interactions between two qutrits [Fig.~\ref{fig:figureone}(d)]. A specific configuration for this tunable coupling is shown in Fig.~\ref{fig:figureone}(e), where nearest-neighbor transmons are coupled via frequency-tunable couplers. Adjusting the frequency of the coupler modifies the strength of the cross-Kerr interaction, which, in combination with single-qutrit operations, enables the realization of efficient conditional phase gates~\cite{Luo2023}. Alternatively, the conditional phase gate can also be dynamically realized by the differential AC Stark shift on two transmon qutrits with static coupling~\cite{Goss2022}.

To prepare the ground state $\ket{\Psi_0}$ in Eq.~\eqref{eq:gs}, we first select a set of representative qutrits on light plaquettes [red dots in Fig.~\ref{fig:figureTwo}(b)] and apply the generalized Hadamard gate to them. The generalized Hadamard gate is defined by the quantum Fourier transform~\cite{Wang2020,Cao2011}:
\begin{equation}\label{eq:hadamard}
    H^{(d)} \ket{j}= \frac{1}{\sqrt{d}}\sum_{i=0}^{d-1} \omega^{ij} \ket{i}.
\end{equation}
In the case of qutrits, the generalized Hadamard gate acts on the state $\ket{0}$ as:
\begin{equation}
   \ket{+} := H^{(3)}\ket{0} = \frac{1}{\sqrt{3}} \left(\ket{0} + \ket{1} +\ket{2}\right),
\end{equation}
which is the common eigenstate of $X$ and $X^{\dagger}$ with eigenvalue $1$. Experimentally, the generalized Hadamard gate can be implemented by decomposing the gate into time evolutions as $H^{(3)}=e^{-iH_dt}e^{-iH_ot}$, where $H_o=\sum_{i < j}m_{ij}\ket{i}\bra{j}+h.c.$ is off-diagonal, and $H_d=\text{diag}(\phi_0,\phi_1,\phi_2)$ is diagonal with complex parameters $m_{ij}$ and real parameters $\phi_i$. The generator $H_o$ is experimentally realized by simultaneously microwave driving transitions between the three pairs of energy levels of the qutrit, while $H_d$ is implemented by shifting the phases of drive fields in the resonant control pulse~\cite{Brien2004}.

After applying the $H^{(3)}$ to each representative qutrit, all the light plaquettes (including the light boundary plaquettes) are prepared in the ground state [Fig.~\ref{fig:figureTwo}(c)]. Next, we select the control qutrit on each dark bulk plaquette [green dots in Fig.~\ref{fig:figureTwo}(b)] and apply $CZ^{\dagger}$, $CZ$, and $CX$ gates to prepare all the dark plaquettes in the ground state [Figs.~\ref{fig:figureTwo}(d) and \ref{fig:figureTwo}(e)]. The controlled two-qutrit gates must be applied layer by layer, ensuring that the state of the representative qutrit remains unchanged until the controlled operations on their respective plaquettes are applied. Ground state preparation on the dark plaquettes does not cause any light plaquette to deviate from the ground state, as their stabilizers commute with each other. In this way, we prepare the entire system in the ground state. The depth of the ground state preparation circuit depends linearly on the size of the lattice ($\sqrt{n}$ in our set up, with $n$ the number of qutrits).

An alternative, more efficient method exists if the system allows for non-destructive measurement for all plaquette states simultaneously. This requires appending an ancillary qutrit to each plaquette. We then apply the generalized Hadamard test. More precisely, we first initialize all the ancillary qutrits in the $\ket{+}$ state, and the total state on the plaquette $p$ is:
\begin{equation}\label{eq:gspI}
    \ket{\Psi_{tot}} = \frac{1}{\sqrt{3}}\left(\ket{\tilde 0}+\ket{\tilde 1}+\ket{\tilde 2}\right)\otimes\ket{\Psi_p},
\end{equation}
where the $\ket{\tilde j}$ denotes the ancillary state, and $\ket{\Psi_p}$ denotes the plaquette state. Then we act $O_p$ conditioned on the ancillary state, transforming the state as
\begin{equation}\label{eq:gspII}
  \tilde CO_p\ket{\Psi_{tot}} = \frac{1}{\sqrt{3}}\left(\ket{\tilde 0}\ket{\Psi_p}+\ket{\tilde 1}O_p\ket{\Psi_p}+\ket{\tilde 2}O_p^{\dagger}\ket{\Psi_p}\right),
\end{equation}
where tilde indicates that the conditional gate $\tilde CO_P$ is controlled by the ancillary qutrit. The gate $\tilde CO_P$ consists of  $\tilde CZ_1$, $\tilde CX_2$, $\tilde CZ_3^{\dagger}$, and $\tilde CX_4^{\dagger}$ gates, where the subscripts denote the order of qutrits in a stabilizer, as defined in Eq.~\eqref{eq:ham}. After applying a generalized Hadamard gate to each ancillary qutrit, we measure all ancillary states at the same time to project each of them onto $\ket{\tilde 0}$. The plaquette state transforms as
\begin{eqnarray}\label{eq:gspIII}
    \tilde P(0) \tilde H^{(3)} \tilde CO_p\ket{\Psi_{tot}} &=& \frac{1}{3}\left(\mathbbm{1}+O_p+O^{\dagger}_p\right)\ket{\Psi_p} \nonumber
    \\
    &=& \Pi_p^{0}\ket{\Psi_p},
\end{eqnarray}
where $\tilde H^{(3)}$ indicates that the generalized Hadamard gate acts on the ancillary qutrit, and $\tilde P(0)$ is a projector that projects the ancillary qutrit onto the $\ket{\tilde 0}$ state. Therefore, after the procedure described in Eqs.~\eqref{eq:gspI},~\eqref{eq:gspII}, and~\eqref{eq:gspIII}, each plaquette is subjected to a ground-state projector and prepared in the ground state. This method can be directly generalized to the more general qudit cases.

Fidelity of the ground state preparation can be obtained by the standard quantum state tomography (QST) method to read out the density matrix~\cite{Bianchetti2010}. With the density matrix, the expectation value $\langle \Pi_p^0 \rangle$ of the ground state projector on each plaquette can be computed.  By repeating the experiment multiple times, one obtains the average value $\overline{\langle \Pi_p^0 \rangle}$ for each plaquette. If $\overline{\langle \Pi_p^0 \rangle}=1$, it indicates perfect preparation of the ground state on the plaquette $p$; if $\overline{\langle \Pi_p^0 \rangle}<1$, it indicates that the plaquette deviates from the ground state. This is the typical method used to characterize ground state preparation in digital simulations with SC qubits, and it is extended to the qutrit case in this work.

\subsection{Creation of parafermions}
With the ground state prepared, we now proceed to create a pair of parafermions. To achieve this, we select two neighboring qutrits and measure them in the $XZ$ basis if they are vertically  adjacent or in the $XZ^{\dagger}$ basis if they are horizontally adjacent. By measurement, we project the qutrits onto the eigenstate corresponding to the eigenvalue $1$ of $XZ$ or $XZ^{\dagger}$: $\frac{1}{\sqrt{3}}(\ket{0}+\overline\omega\ket{1}+\ket{2})$ for $XZ$, and $\frac{1}{\sqrt{3}}(\ket{0}+\omega\ket{1}+\ket{2})$ for $XZ^{\dagger}$. Since $XZ$ and $XZ^{\dagger}$ are creation operators of $e\overline m-\overline em$ excitation pairs, this projection projects the qutrits in the coherent state of these creation operators, effectively condensing the $e\overline m$ and $\overline em$ excitations in the local region [the light blue region in Fig.~\ref{fig:figureTwo}(f)]. Because the two qutrits are projected separately, they are not directly entangled, which effectively removes the edge between them, thereby creating a pair of parafermions in the condensate.

A method arises when natural holes appear in the lattice [Fig.~\ref{fig:figureTwo}(g)], potentially resulting from imperfections in the geometric arrangement of qutrits due to fabrication constraints. Exploiting this imperfection, a pair of parafermions can be created by preparing the ground state on the imperfect lattice. For the surface containing the hole, however,  additional controlled gates, such as the controlled-$XZ$ gate, are required to enforce the ground-state condition. An example featuring a minimal hole is illustrated in Fig.~\ref{fig:figureTwo}(g).

\begin{figure*}
    \centering
    \includegraphics[scale=0.24]{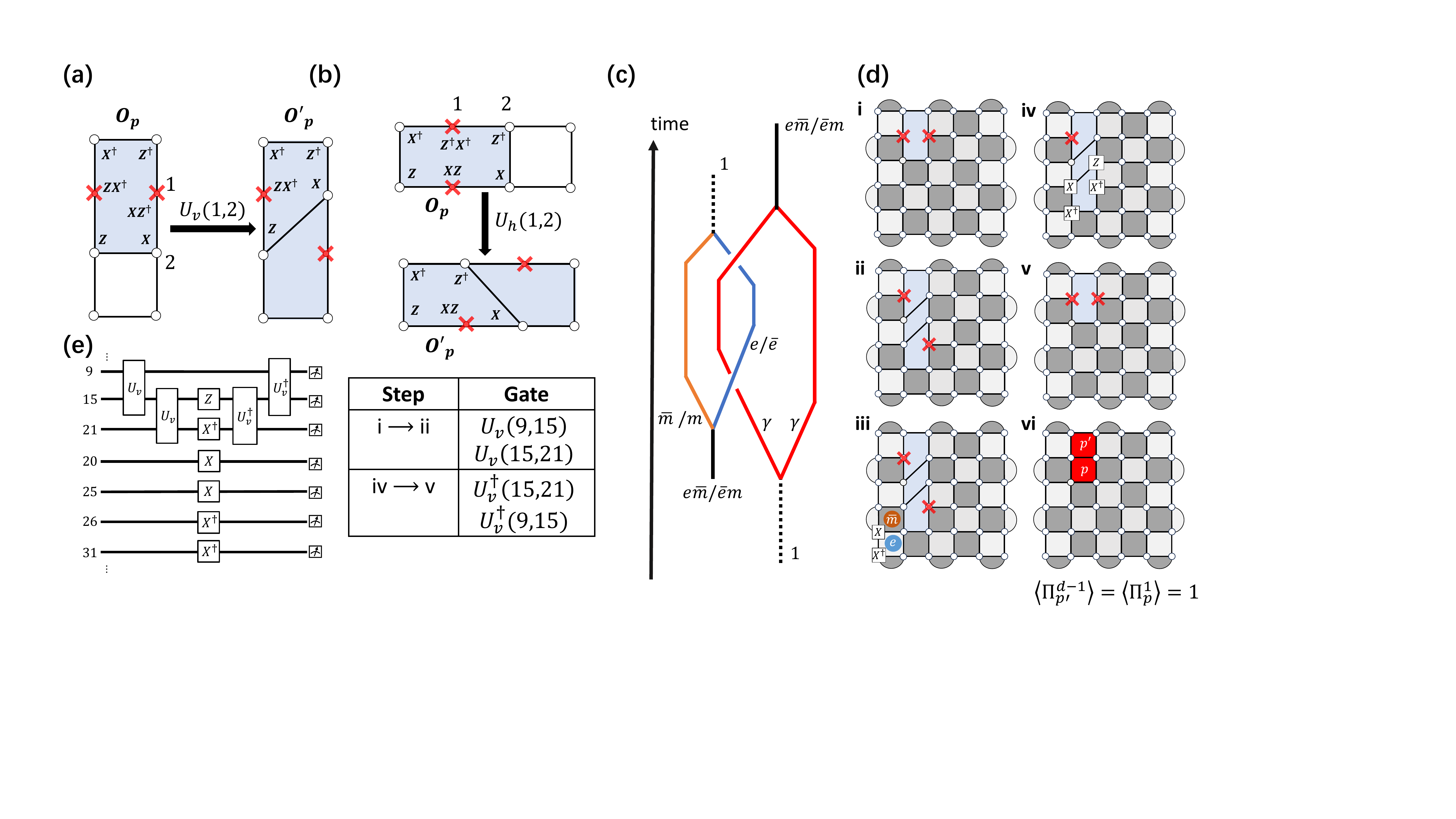}
    \caption{Generalized code deformation and the fusion of parafermions. (a) Code deformation that moves the parafermion vertically. (b) Code deformation that moves the parafermion horizontally. (c) Illustration of the creation of a composite excitation and braiding an $e$ or $\overline e$ excitation with one parafermion. After the braiding, the excitations fuse to the trivial topological charge. (d) Explicit procedure for realizing the fusion of parafermions. Observables are characterized by $\langle\Pi_{p'}^{d-1}\rangle=\langle\Pi_{p}^{1}\rangle=1$. (e) Quantum circuit and two-qutrit gates used in the procedure.}
    \label{fig:figureThree}
\end{figure*}

\section{FUSION OPERATIONs OF THE GENERATED PARAFERMIONS }\label{sec:fusion}

In this section, we propose a scheme to realize parafermion fusion. While the previous section presented a method to create parafermion pairs via engineered dislocations, the adjacent creation of parafermions hinders the study of fusion rules of individual parafermion, as nearby parafermion can influence the fusion result. To address this, we generalize the code deformation method to qudit systems and derive operators that enable spatial separation of parafermions. This generalization is technically nontrivial due to the non-Hermitian nature of the stabilizers, requiring careful treatment of both the stabilizer and its Hermitian conjugate. Physically, this complexity is intrinsically linked to the non-self-duality of anyons and the nonlinearity of parafermion operator transformations—a connection we leave to future work. The resulting code deformation protocol allows both horizontal and vertical movement of parafermions on the surface, offering a versatile tool for their manipulation and enabling further operations such as braiding (see Sec.~\ref{sec:braiding}). Given its significance and fundamental role in this study, we first introduce the generalized code deformation method before combining it with string operators to establish a protocol for realizing the fusion of parafermions.

\subsection{Generalized code deformation}
Code deformation is a method used to convert one error-correcting code into another by modifying the stabilizers, enabling fault-tolerant logical gate operations of surface codes~\cite{Vuillot2019,Bombin2011}. Beyond quantum information science, code deformation also finds applications in condensed matter physics. Topological surface codes, such as the toric code, are rooted in topological quantum field theory, which provides the general framework of topological orders. In this context, code deformation involves altering the topological structure of the field theory, such as engineering topological defects, to enrich the properties of topological orders~\cite{Lensky2023}. While the method has been extensively studied in qubit systems, its application to qudit systems remains relatively unexplored. In this subsection, we generalize the code deformation method to the qudit system. This generalization enables the derivation of unitary operators that facilitate the spatial movement of parafermions on the surface. Our work extends the applicability of code deformation to a broader class of qudit surface codes.

Given a qudit stabilizer $O_p$ on the plaquette $p$ and Hamiltonian $H=-\sum_pO_p+h.c.$, the local ground state $\ket{\psi}$ satisfies $O_p\ket{\psi}=O_p^{\dagger}\ket{\psi}=\ket{\psi}$. After a code deformation, such as a change in the geometry of the lattice, the stabilize changes to $O'_p$, and the corresponding local ground state is $\ket{\psi'}$. The new local ground state is related to the original one by a unitary operator $U$: $\ket{\psi'}=U\ket{\psi}$. For qutrits,  the unitary operator is given by
\begin{equation}\label{eq:gcd}
    U = \frac{1}{\sqrt{3}}\left(\mathbbm{1}+O'^{\dagger}_pO_p+O'_pO_p^{\dagger}\right).
\end{equation}
For our purposes, we study the code deformations depicted in Fig.~\ref{fig:figureThree}(a) and Fig.~\ref{fig:figureThree}(b), where the former moves the parafermion vertically and the latter moves the parafermon horizontally. The associated unitary operator for the vertical move is:
\begin{equation}
    U_v(i,j) = \frac{1}{\sqrt{3}}\left(\mathbbm{1}+\omega Z_i^{\dagger} X_j+Z_i X_j^{\dagger}\right),
\end{equation}
and for the horzontal move:
\begin{equation}
    U_h(i,j) = \frac{1}{\sqrt{3}}\left(\mathbbm{1}+\omega Z_j X_i+Z_j^{\dagger}X_i^{\dagger}\right),
\end{equation}
where $i$ and $j$ denote the initial and final positions of the parafermion, respectively. Both $U_v$ and $U_h$ are two-qutrit operators, and their matrices in the basis $\{\ket{00},\ket{01},\ket{02},\ket{10},\ket{11},\ket{12},\ket{20},\ket{21},\ket{22}\}$ are:

\begin{equation}\label{eq:uvm}
U_v = \frac{1}{\sqrt{3}}\begin{pmatrix}1 & \omega &1 &0 & 0&0 &0 &0&0\\
1 & 1 & \omega &0 & 0&0 &0 &0&0\\
\omega & 1 & 1 &0 & 0&0 &0 &0&0\\
0 & 0 & 0 &1 & 1& \omega &0 &0&0\\
0 & 0 & 0 &\omega & 1&1 &0 &0&0\\
0 & 0 & 0 &1 & \omega&1 &0 &0&0\\
0 & 0 & 0 &0 & 0&0 &1 &\overline \omega&\overline \omega\\
0 & 0 & 0 &0 & 0&0 &\overline \omega &1&\overline \omega\\
0 & 0 & 0 &0 & 0&0 &\overline \omega &\overline \omega&1\\
\end{pmatrix},
\end{equation}
and
\begin{equation}\label{eq:uhm}
U_h = \frac{1}{\sqrt{3}}\begin{pmatrix}
 1 & 0 & 0 & \omega & 0 & 0 & 1 & 0 & 0 \\
 0 & 1 & 0 & 0 & \overline \omega & 0 & 0 &
   \overline \omega & 0 \\
 0 & 0 & 1 & 0 & 0 & 1& 0 & 0 & \omega  \\
 1 & 0 & 0 & 1 & 0 & 0 & \omega & 0 & 0 \\
 0 & \overline \omega & 0 & 0 & 1 & 0 & 0 &
   \overline \omega & 0 \\
 0 & 0 & \omega & 0 & 0 & 1 & 0 & 0 & 1 \\
 \omega & 0 & 0 & 1 & 0 & 0 & 1 & 0 & 0 \\
 0 & \overline \omega & 0 & 0 & \overline \omega & 0 & 0 & 1 & 0 \\
 0 & 0 & 1 & 0 & 0 & \omega & 0 & 0 & 1 \\
\end{pmatrix}.
\end{equation}
Before proceeding to the realization of the operators $U_h$ and  $U_v$ using quantum circuits, we briefly review the code deformation method used to move Majorana modes in stabilizer codes with SC qubits~\cite{Xu2023,Andersen2023} and compare it with our generalization. In the qubit case, both the original and deformed stabilizers \( S_{\text{old}} \) and \( S_{\text{new}} \) are Hermitian. This allows a straightforward construction of a unitary operator that maps an eigenstate of \( S_{\text{old}} \) with eigenvalue \( +1 \) to an eigenstate of \( S_{\text{new}} \) with the same eigenvalue. Specifically, since \( S_{\text{old}} \) and \( S_{\text{new}} \) anticommute, the product \( i S_{\text{new}} S_{\text{old}} \) is Hermitian, motivating the construction of $e^{\frac{\pi}{4} S_{\text{new}}S_{\text{old}}}$. Applying this operator achieves the desired state transformation. In the qudit case, however, the stabilizers are no longer Hermitian, and this strategy breaks down. The lack of Hermiticity introduces significant complications in constructing the unitary operator defined in Eq.~\eqref{eq:gcd}, making the generalization highly non-trivial.

\begin{figure}
    \centering
    \includegraphics[scale=0.3]{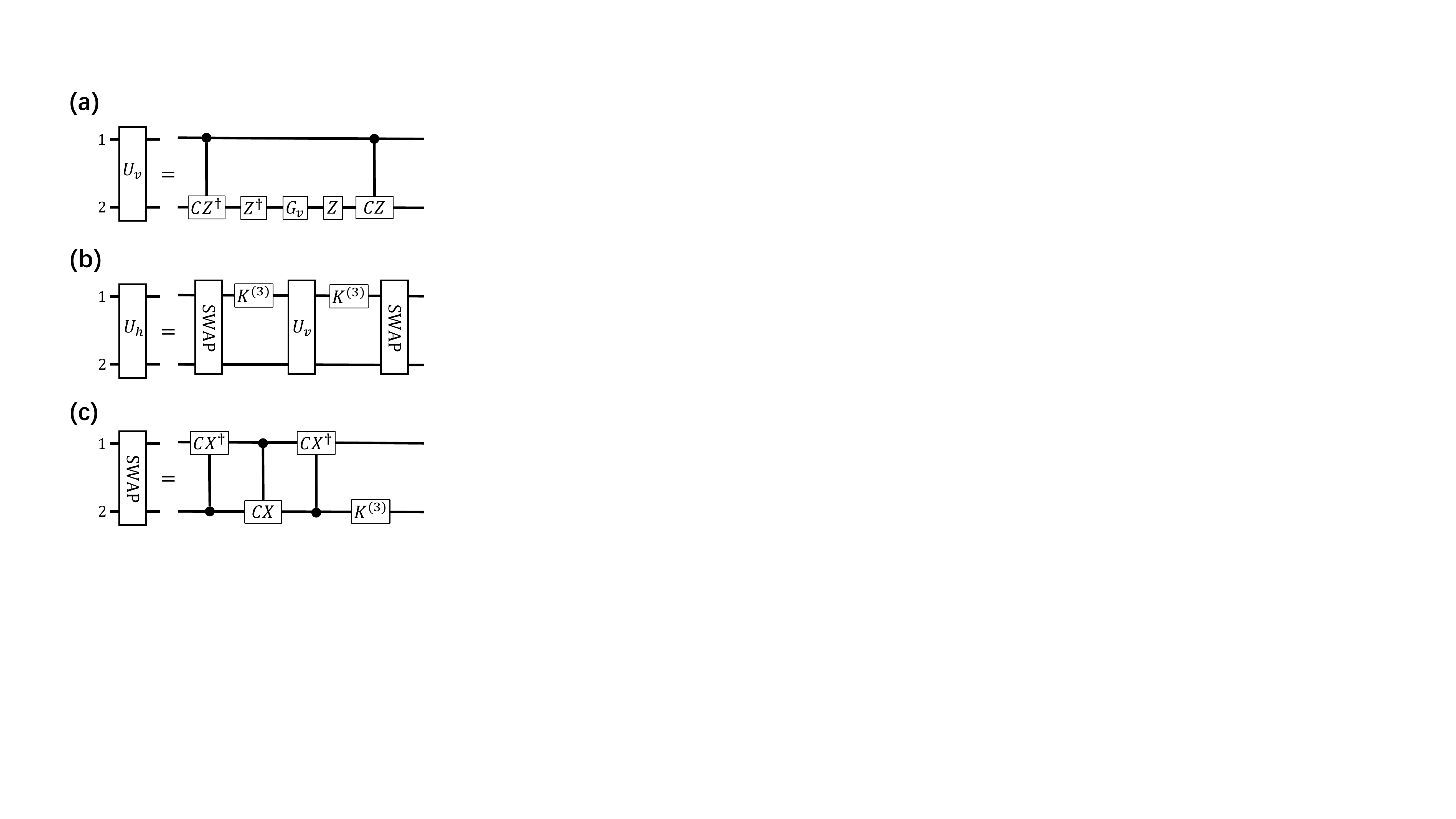}
    \caption{(a) Quantum circuit that realizes the $U_v$ operator. (b) Quantum circuit that transforms the $U_v$ operator into the $U_h$ operator. (b) The qutrit SWAP circuit. }
    \label{fig:SWAP}
\end{figure}

The two-qutrit operators $U_h$ and $U_v$ belong to the Clifford group, meaning they can be realized with a universal set of qutrit gates. The specific circuit for realizing $U_v$ operator is shown in Fig.~\ref{fig:SWAP}(a), where $G_v$ is a single-qutrit operator decomposable into rotations within the $\{\ket{0},\ket{1}\}$ and $\{\ket{1},\ket{2}\}$ subspaces, along with diagonal phase matrices:
\begin{small}
\begin{eqnarray}
G_v=
&\begin{pmatrix}
 1 & 0 & 0 \\
 0 & e^{\frac{i\pi}{3}} & 0 \\
 0 & 0 & e^{\frac{i\pi}{3}}
\end{pmatrix}
\begin{pmatrix}
 1 & 0 & 0 \\
 0 & \text{cos}(\theta_1) & -\text{sin}(\theta_1) \\
 0 & \text{sin}(\theta_1) & \text{cos}(\theta_1)
\end{pmatrix} \nonumber
\\
&\begin{pmatrix}
 \text{cos}(\theta_2) & -\text{sin}(\theta_2) & 0 \\
 \text{sin}(\theta_2) & \text{cos}(\theta_2) & 0\\
 0 & 0 & 1
\end{pmatrix}
\begin{pmatrix}
 1 & 0 & 0 \\
 0 & e^{\frac{i4\pi}{3}} & 0 \\
 0 & 0 & e^{\frac{i5\pi}{6}}
\end{pmatrix} \nonumber
\\
&\begin{pmatrix}
 1 & 0 & 0 \\
 0 & \text{cos}(\theta_1) & -\text{sin}(\theta_1) \\
 0 & \text{sin}(\theta_1) & \text{cos}(\theta_1)
\end{pmatrix}
\begin{pmatrix}
 1 & 0 & 0 \\
 0 & 1& 0 \\
 0 & 0 & -1
\end{pmatrix},
\end{eqnarray}
\end{small}
where $\theta_1=\frac{\pi}{4}$ and $\theta_2\approx 1.696\pi$. These rotation and diagonal matrices can be experimentally implemented using microwave drives for level transitions and phase shifts of drive pulses~\cite{Yurtalan2020,Morvan2021}. The circuit implementation for $U_v$ is not necessarily unique and optimization algorithms can enhance its efficiency~\cite{Dhruv2024}. Machine learning methods may offer further  improvements. Once $U_v$ is implemented, the operator $U_h$ can be obtained using the circuit in Fig.~\ref{fig:SWAP}(b),  where the qutrit SWAP gate is implemented using the circuit in Fig.~\ref{fig:SWAP}(c). Here, the $K^{(3)}$ gate is defined by $K^{(3)}\ket{j} = \ket{3-j}$.

\subsection{Realization of fusion operations}\label{ssec:fusionoperations}

\begin{figure*}
    \centering
    \includegraphics[scale=0.24]{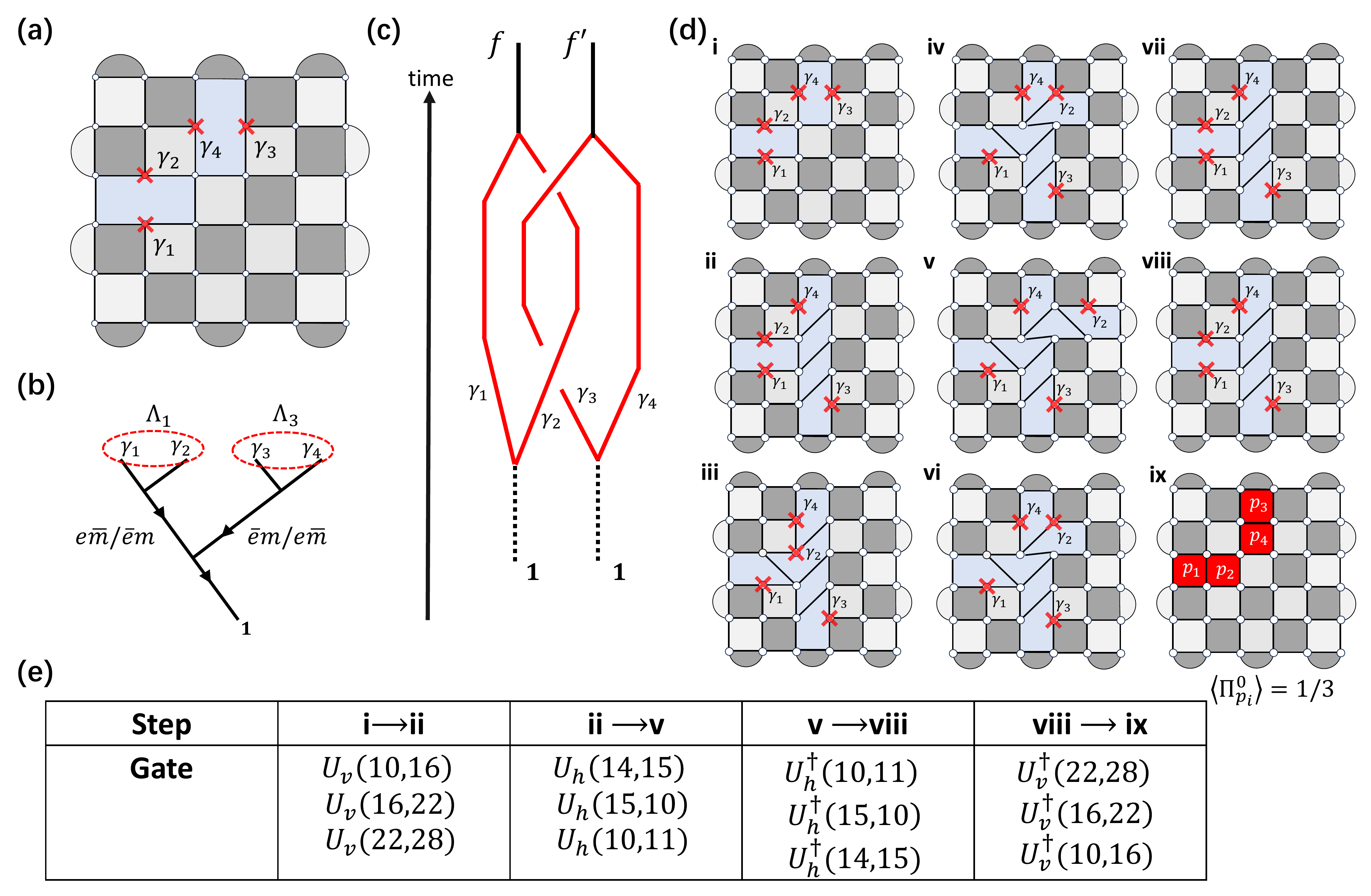}
    \caption{Braiding of parafermion modes. (a) The $\mathbb{Z}_3$ plaquette model with four parafermions denoted by $\gamma_1$, $\gamma_2$, $\gamma_3$, and $\gamma_4$ labeld by the red crosses. (b) The fusion tree of parafermions. (c) Illustration of the creation of two pairs of parafermions from the ground state and the full braiding between two parafermions. After the braiding, they fuse to two composite excitations denoted by $f$ and $f'$, respectively. (d) Explicit procedure for the full braiding of parafermions $\gamma_2$ and $\gamma_3$. (e) Two-qutrit gates used in the procedure.}
    \label{fig:figureFour}
\end{figure*}
We now present the protocol for realizing the fusion of parafermions. The characteristic fusion rules for parafermions are:
\begin{equation}\label{eq:qutritfusion}
    \gamma \times \gamma = 1 + e\overline m + \overline e m,
\end{equation}
which is a special case of Eq.~\eqref{eq:fusionrule}. This rule indicates that the parafermion provides the fusion channels that identify the $e\overline m$ and $\overline e m$ excitations with the trivial topological charge $1$. The related fusion rule is:
\begin{equation}\label{eq:absorb}
    \gamma \times e\overline m/\overline em = 1,
\end{equation}
which shows that a parafermion can absorb either the $e\overline m$ or $\overline e m$ excitations. This can be achieved by braiding an $e$ or $\overline e$ excitation with a parafermion [Fig.~\ref{fig:figureThree}(c)].

In Fig.~\ref{fig:figureThree}(d), we explicitly illustrate the procedure to perform the fusions in Eq.~\eqref{eq:absorb}. The protocol begins by measuring two horizontally adjacent qutrits in the $XZ^{\dagger}$ basis and projecting them onto the eigenvalue-$1$ eigenstate, thereby creating a pair of parafermions. Since parafermions are created from the ground state, their initial state is in a fixed fusion channel: $\ket{\gamma,\gamma;1}$. We then separate the two parafermions by moving one downward using the operator $U_v$. Next, an $e\overline m$ dyon excitation is created by applying two Pauli operators. The $e$ excitation is then braided around one parafermion using a string operator, which consists of a sequence of Pauli operators. After braiding, the $e$ excitation is transformed into the $m$ excitation and proceeds to annihilate the $\overline m$ excitation. Once this process is completed, we move the parafermion back to a neighboring position and measure the expectation values $\langle \Pi^{d-1}_{p'}\rangle$ and $\langle \Pi^{1}_{p}\rangle$ by the QST method [the (vi) step in Fig.~\ref{fig:figureThree}(d)]. Before fusion, both $\langle \Pi^{d-1}_{p'}\rangle$ and $\langle \Pi^{1}_{p}\rangle$ vanish due to the initial fusion channel $\ket{\gamma,\gamma;1}$. After fusion, the parafermon absorbs the $e\overline m$ excitation, resulting in the predicted expectation values $\overline{\langle \Pi^{d-1}_{p'}\rangle}=1$ and $\overline{\langle \Pi^{1}_{p}\rangle}=1$, averaged over multiple experimental realizations. A similar procedure can be applied to create and braid the $\overline e m$ excitation, providing an experimentally accessible observable to realize the fusion rules in Eq.~\eqref{eq:absorb}. In particular, reversing steps (iii-iv) in Fig.~\ref{fig:figureThree}(d) implements the fusion rule in Eq.~\eqref{eq:qutritfusion}. In Fig.~\ref{fig:figureThree}(e), we demonstrate the corresponding circuit and two-qutrit operators used in the procedure.

\section{Braiding of parafermion modes}\label{sec:braiding}

After realizing the fusion of parafermions, we now turn to the braiding of them.
In Sec.~\ref{sec:fusion}, we introduce the generalized code deformation method, which provides two unitary operators that move parafermions vertically and horizontally [Figs.~\ref{fig:figureThree}(a) and \ref{fig:figureThree}(b)]. These operations enable the full braiding of parafermions. To characterize the braiding statistics, we define the computational space basis encoded by parafermions. Compared to Majorana modes, the computational space of $\mathbb{Z}_3$ parafermions naturally encodes logical qutrits rather than qubits. This distinction arises from the difference in fusion rules between parafermions and Majorana mdoes, with the former having more fusion channels. Within this larger computational space, parafermions exhibit more intricate braiding statistics, making them harder to measure. We identify the expectation value of local ground-state projectors as an observable for characterizing braiding results.

To encode information, a minimum number of four parafermions is required [Fig.~\ref{fig:figureFour}(a)]. In this configuration, there are only two parity operators, $\Lambda_1=\overline{ \omega}\gamma_1\gamma_2^{\dagger}$ and $\Lambda_3=\overline{ \omega}\gamma_3\gamma_4^{\dagger}$ with $\gamma$ the parafermion operator (see Appendix.~\ref{sec:algebra}). For convenience, the computational space is restricted to the two-dimensional subspace spanned by $\Lambda_1\Lambda_3=1$. With this restriction, four parafermions are sufficient to encode a logical qutrit. To find the basis of the computational space, we diagonalize the parity operator $\Lambda_i$ and obtain the eigenstates $\{\ket{0}_{\Lambda_i}, \ket{1}_{\Lambda_i}, \ket{2}_{\Lambda_i}\}$. The logical qutrit state is defined by:
\begin{equation}\label{eq:qutritbasis}
    \ket{\overline j} = \ket{j}_{\Lambda_1}\otimes\ket{3-j}_{\Lambda_3}.
\end{equation}
On the other hand, physically, the basis of the computational space is defined by the fusion tree of the four parafermions [Fig.~\ref{fig:figureFour}(b)]. The specific transformation between the basis defined by the eigenstates of the parity operators and the fusion basis $\ket{\gamma_1,\gamma_2;1/e\overline m/\overline e m}\otimes\ket{\gamma_3,\gamma_4;1/\overline em/e\overline m}$ involves selecting two base points on the surface and constructing the parafermion operators using string operators connecting these base points, with carefully selected arrows to satisfy commutation relations Eq.~\eqref{eq:algebra}~\cite{Bombin2010,Xu2023}. Notably, the representation of parafermion operators via string operators is reminiscent of the Jordan-Wigner transformation, highlighting their non-locality. This non-local nature stems from the defining non-local commutation relations of parafermions Eq.~\eqref{eq:algebra}. Since these string operators share common endpoints, the parity operators Eq.~\eqref{eq:parity} function as loop operators, measuring the fusion results of parafermions enclosed within the loop. Consequently, the logical qutrit is globally defined when the parafermions are sufficiently separated. This is also evident in the fusion operations discussed in Sec.~\ref{ssec:fusionoperations}, where transforming the logical qutrit state involves creating a composite excitation $e\overline{m}$ adjacent to the boundary and braiding either $e$ or $\overline{m}$ around the parafermion. When parafermions are engineered at the center of the system, such fusion operations require a string operator connecting the boundary to the parafermion defect and a loop operator encircling the defect. These operations consist of a set of local operators whose number, i.e., the code distance, increases linearly with system size, thereby ensuring topological protection of the logical qutrit state. In this study, we adopt a more direct approach to measure the braiding statistics of parafermions without relying on the explicit construction of parity operators and the detailed relationship between the fusion basis and the logical qutrit basis.

In the logical qutrit basis $\{\ket{\overline j}\}$, the half-braiding operator $U_2$ that exchanges $\gamma_2$ and $\gamma_3$ is an off-diagonal matrix, exhibiting the non-Abelian statistics. In a specific representation, the matrix elements of $U_2$ are given by: $\bra{\overline i}U_2\ket{\overline j} = \frac{1}{\sqrt{3}} c_{i-j},$ where $c_j = \omega^{\frac{j(j+5)}{2}}$ (see Appendix.~\ref{sec:algebra}). The full braiding matrix in the logical qutrit basis is:
\begin{equation}\label{eq:brmatrix}
    U_2^2 = \frac{1}{\sqrt{3}}\begin{pmatrix}
 i & i & \frac{2+\overline \omega}{\sqrt{3}} \\
 \frac{2+\overline \omega}{\sqrt{3}} & i & i \\
 i & \frac{2+\overline \omega}{\sqrt{3}} & i \\
\end{pmatrix},
\end{equation}
and acts on the logical qutrit state $\ket{\overline 0}$ as:
\begin{equation}\label{eq:braidng}
    U_2^2\ket{\overline 0} = \frac{i}{\sqrt{3}}\ket{\overline 0}+\frac{2+\overline\omega}{3}\ket{\overline 1} +\frac{i}{\sqrt{3}}\ket{\overline 2}.
\end{equation}
The key point is that, regardless of how the logical states are identified in terms of the fusion basis, the fusion state corresponding to the trivial charge, $\ket{\gamma_1,\gamma_2;1}\otimes \ket{\gamma_3,\gamma_4;1}$, can always be identified with the state $\ket{\overline{0}}$. This follows from the physical insight that a state with a total trivial topological charge has parity $1$. Since the parafermions are created from the ground state, their initial state is $\ket{\gamma_1,\gamma_2;1}\otimes \ket{\gamma_3,\gamma_4;1}\equiv\ket{\overline 0}$. After a full braiding of $\gamma_2$ and $\gamma_3$, the state transforms as Eq.~\eqref{eq:braidng}.

We now move on to the explicit procedure. In Fig.~\ref{fig:figureFour}(d), we show the implementation of full braiding of parafermions. The process begins by creating four parafermion $\gamma_1$, $\gamma_2$, $\gamma_3$, and $\gamma_4$ from the ground state, arranged in the configuration shown in Fig.~\ref{fig:figureFour}(a). In the ground state, measuring $\langle \Pi_{p_i}^0 \rangle$ in these plaquettes yields $\langle \Pi_{p_i}^0 \rangle = 1$.
We then move $\gamma_3$ to the bottom, followed by moving $\gamma_2$ to the right. After these steps, we move $\gamma_2$ back to its original position and finally return $\gamma_3$ to its starting point. After braiding, the parafermions $\gamma_1$ and $\gamma_2$ fuse to form a composite excitation, denoted by $f$, which consists of the components $1$, $\overline e m$, and $e \overline m$. Similarly, parafermions $\gamma_3$ and $\gamma_4$ fuse into another composite excitation $f'$ [Fig.~\ref{fig:figureFour}(c)]. In general, $f$ and $f'$ are distinct, but they are related by an electric-magnetic transformation, where $e \leftrightarrow m$. According to Eq.~\eqref{eq:braidng}, the probability of measuring the topologically trivial charge $1$ from the composite excitation $f$ or $f'$ is always $\frac{1}{3}$. Therefore, when measuring the expectation value $\langle \Pi_{p_i}^0 \rangle$ using the QST method on the plaquettes that support the dislocation, the averaged result over multiple repeated experiments is given by  $\overline{\langle \Pi_{p_i}^0 \rangle} = \frac{1}{3}$. This provides an experimental observable for the braiding of parafermions. In Fig.~\ref{fig:figureFour}(e), we show two-qutrit gates used in the procedure.

\section{Experimental parameters}\label{sec:experpara}
In this section, we briefly discuss typical parameters for SC circuits relevant to the implementation of our scheme, focusing on the case of $d=3$. A key parameter in SC circuits is the fidelity of single- and two-qutrit gates. Recent experiments have demonstrated the generalized Hadamard gate for qutrits with a high fidelity of $0.992$~\cite{Yurtalan2020}. Implementing high-fidelity two-qutrit gates is more challenging. A recent study reported fidelities of $0.973$ and $0.952$ for the realization of $CZ^{\dagger}$ and $CZ$ gates of qutrits, respectively~\cite{Goss2022}. Since single-qutrit gate fidelities are considerably higher, we neglect their contribution to overall errors in our analysis. Our ground state preparation protocol requires three two-qutrit gates per dark plaquette. To achieve a total fidelity of $0.90$ for ground state preparation, the fidelity of each two-qutrit gate needs to exceed $0.965$. This value is already within reach of current experimental platforms. Achieving even higher fidelities would require further advances in SC technology, marking an important direction for future research. Other important parameters include the relaxation time  $T^{ij}_1$ and the dephasing time $T^{ij}_2$ between different qutrit levels ($ij\in\{01,12,02\}$). These times must be sufficiently long to allow for fusion, braiding, and measurement procedures. Recent experiments on transmon qutrits report typical values of both $T^{ij}_1$ and $T^{ij}_2$ around $10\mu s$ or higher~\cite{Luo2023,Goss2022}. Considering that single-qutrit gate durations are on the order of  $10-100 ns$ and two-qutrit gates take about $100-800 ns$, these coherence times suggest that ground state preparation and subsequent parafermion manipulations are feasible with near-term SC technologies.~\cite{Luo2023}.

In the main text, we use a $6\times6$ qutrit array for illustrative purposes, though this is not the minimal configuration required by our scheme. For studying fusion rules, a $3\times4$ array suffices, while braiding operations can be demonstrated with a $4\times4$ setup. Reducing the number of qutrits per row or column lowers the total number of parafermion movement operations by two, saving hundreds to thousands of nanoseconds in operation time. This reduction leads to a significant improvement in both the success rate and the overall fidelity of the experiment. The $6\times6$ configuration, however, enables more advanced protocols, such as encoding two logical qutrits using eight parafermions and performing their braiding operations. Moreover, by designing quasi-one-dimensional layouts optimized for braiding, it is possible to further minimize the required number of qutrits.

\section{Discussion and conclusion}\label{sec:outlook}
Before concluding, we discuss potential applications in high-dimensional quantum computing. Recent developments highlight several advantages over qudit-based quantum information processing~\cite{Campbell2014} and flexible simulation of quantum dynamics with qudits~\cite{Matthew2009}. Qudit versions of various quantum algorithms have already been developed~\cite{Nguyen2019,Ivanov2012,Cao2011}. High-dimensional quantum computing based on non-Abelian parafermions is inherently fault-tolerant due to topological protection. While realizing parafermions in physical materials remains a great challenge, our quantum simulation scheme offers a promising approach to high-dimensional topological quantum computing in the near future. One exciting direction is to implement qudit quantum algorithms using parafermions braidings. Another challenge is to implement the topologically protected $CX$ gate in quantum simulators, involving eight parafermions and twelve half-braidings~\cite{Hutter2016}.

In summary, we have proposed in this work a novel experimental scheme for realizing and manipulating non-Abelian parafermions in SC circuits, which is also applicable to other quantum simulation and quantum computing platforms, including the trapped ions and neutral atom arrays which develop fast recently. We have demonstrated efficient protocols for ground state preparation and parafermion creation, and proposed a generalized code deformation approach to achieve the fusion rules and braiding statistics through experimental observables. This work with experimental feasibility paves the way for the quantum simulation of parafermions and their non-Abelian statistics in SC circuits.

\section*{Acknowledgments}

This work was supported by National Key Research and Development Program of China (2021YFA1400900), the National Natural Science Foundation of China (Grants No.~12425401 and No.~12261160368),  the Innovation Program for Quantum Science and Technology (Grant No.~2021ZD0302000), and by Shanghai Municipal Science and Technology Major Project (Grant No.~2019SHZDZX01).

\appendix
\renewcommand{\appendixname}{APPENDIX}

\section{Elementary excitations of the model}\label{sec:elementarryex}
In this section, we investigate the elementary excitations of the model Hamiltonian Eq.~\eqref{eq:ham}. The $\mathbbm{Z}_d$ plaquette model is an exactly solvable Hamiltonian model, of which the spectrum includes a series of elementary excitations. To characterize them, we define the following projectors:
\begin{equation}
\Pi_p^{\lambda} = \frac{1}{d} \sum_{k = 0}^{d-1} \omega^{\lambda(d-k)} O_p^k,
\end{equation}
where $\lambda = 0,1,..., d-1$. These projectors are observables $[\Pi_p^{\lambda},H] = 0$ and satisfy
\begin{eqnarray}
\sum_{\lambda} \Pi^{\lambda}_p &=& \mathbbm{1}, \, \text{(completeness)}
\\
\Pi^{\lambda}_p\Pi^{\lambda'}_p &=& \delta_{\lambda\lambda'}\Pi^{\lambda}_p. \, \text{(orthogonality)}
\end{eqnarray}
Therefore, their eigenvalues define a set of topological charges $\{\pi_p^0,\pi_p^{1},...,\pi_p^{\lambda},...,\pi_p^{d-1}\}$ for each plaquette $p$. In particular, the ground states have the topological charge $\{1,0,...,0\}$.

Among the excitation states, we identify that states with topological charge $\{0, 1, \dots, 0\}$ on a given light (dark) plaquette $p$ correspond to a charge (flux) excitation $e$ ($m$) on $p$. Additional quantum numbers $\pi_p^{\lambda} = 1$ indicate the presence of a charge (flux) excitation $e^{\lambda}$ ($m^{\lambda}$) on the light (dark) plaquette $p$. For convenience, we define $\overline{e} \equiv e^{d-1}$ and $\overline{m} \equiv m^{d-1}$. These elementary excitations can be created in pairs and moved via string operators. More general composite excitations are formed from these elementary excitations.

\section{Algebraic theory of parafermions and braiding operator}\label{sec:algebra}
In this section, we introduce the algebraic theory of $\mathbb{Z}_3$ parafermions. For a general $d$, we refer to Ref.~\cite{Hutter2016}. The algebraic relations for $\mathbb{Z}_3$ parafermions are given by:
\begin{equation}\label{eq:algebra}
    \gamma_i\gamma_j = \omega^{\text{sgn}(j-i)} \gamma_j\gamma_i, \quad \gamma_i^3 = \mathbbm{1},
\end{equation}
for an ordered set $\{i\}$ with $\omega = e^{\frac{2\pi i}{3}}$. To process quantum information, we need to consider $N$ pairs of parafermions and unitary representations of the braid group $B_{2N}$. It turns out that there are six representations of the braid group in terms of the parity operators $\Lambda_i$, which are defined as:
\begin{equation}\label{eq:parity}
    \Lambda_i = \overline \omega \gamma_i \gamma_{i+1}^{\dagger},
\end{equation}
where $\overline{\omega}$ denotes the complex conjugate of $\omega$. These operators satisfy the following relations:
\begin{eqnarray}
\Lambda_i \Lambda_j=\Lambda_j \Lambda_i, & \text { if }|i-j|>1 \\
\Lambda_i \Lambda_j=\omega^{\operatorname{sgn}(j-i)} \Lambda_j \Lambda_i. & \text { if }|i-j|=1
\end{eqnarray}
The half-braiding operator that exchanges $\gamma_i$ and $\gamma_{i+1}$ is given by:
\begin{equation}
   U_i=\frac{1}{\sqrt{3}} \sum_{m =0}^2 c_m\left(\Lambda_i\right)^m.
\end{equation}
These operators must satisfy the Yang-Baxter equation:
\begin{equation}
    U_i U_{i+1} U_i=U_{i+1} U_i U_{i+1}.
\end{equation}
Solving the equation yields a family of six representations, which differ by the coefficients $c_m$:
\begin{equation}
    c_m=\omega^{\pm\frac{m(m+2r+3)}{2}}, \, r=0,1,2.
\end{equation}
In this paper, we focus on a specific representation where $c_m=\omega^{\frac{m(m+5)}{2}}$, and the corresponding operator transformations under half-braiding are:
\begin{eqnarray}
U_i\gamma_iU_i^{\dagger}  &=& \overline\omega \gamma_{i+1}, \nonumber
\\
U_i\gamma_{i+1}U_i^{\dagger} &=& \gamma_i^{\dagger}\left(\gamma_{i+1}\right)^2.
\end{eqnarray}
With this representation, the half-braiding operator $U_i$ becomes:
\begin{equation}
    U_i = \frac{1}{\sqrt{3}}\left(1+\overline\omega\gamma_i\gamma_{i+1}^{\dagger}+\overline\omega\gamma_{i+1}\gamma_i^{\dagger}\right),
\end{equation}
and the full braiding operator is $U_i^2$.

For a pair of $\mathbb{Z}_3$ parafermions, the computational basis, which is subject to the constraint $\Lambda_1 \Lambda_3 = 1$, is given by the eigenstates of the parity operators $\Lambda_i$:
\begin{equation}
 \Lambda_i\ket{m}_{\Lambda_i}=m\ket{m}_{\Lambda_i}.
\end{equation}
We denote the basis as $\ket{\overline{k}} = \ket{k}_{\Lambda_1} \otimes \ket{3-k}_{\Lambda_3}$ with $k = 0, 1, 2$. In this basis, the matrix representation of the half-braiding operator $U_1$ is diagonal:
\begin{equation}
    U_1=\sum_{k=0}^2 \tilde c_k\ket{\overline k}\bra{\overline{k}},
\end{equation}
where $\tilde c_k=\frac{1}{\sqrt{3}}\sum_{j=0}^2\omega^{jk}c_j$ is the Fourier transform of $c_m$. The half-braiding operator $U_2$, however, is off-diagonal:
\begin{eqnarray}
    \langle\overline k|U_{2}| \overline l\rangle &=& \langle\overline k|F^{\dagger}U_{1}F| \overline l\rangle \nonumber
    \\
    &=& \frac{1}{\sqrt{3}}c_{k-l} = \frac{1}{\sqrt{3}}\omega^{\frac{(k-l)(k-l+5)}{2}},
\end{eqnarray}
where $F$ is the Fourier transformation operator $F=\frac{1}{\sqrt{3}} \sum_{k, m =0}^2 \omega^{k m}|\overline k\rangle\left\langle\left. \overline l\right|\right.$. Then, we obtain the full braiding matrix $U_2^2$ as shown in Eq.~\eqref{eq:brmatrix}.

%\section*{}

%\textbf{Data Availability.}--The data that support the findings of this study are available from the corresponding author upon reasonable request.

\nocite{*}
%\bibliography{digitalsimparafermionprr}
%\bibliographystyle{apsrev4-2}
% Produces the bibliography via BibTeX.


\end{document}